\documentclass[aps,prl,floats,twocolumn,floatfix,showpacs]{revtex4}
\usepackage{epsf,graphicx}
\usepackage{amssymb}
\usepackage{amsmath}
\usepackage{eepic}

\newlength{\piclen}
\begin{document}

\title{Doped Mott insulator as the origin of heavy Fermion behavior in LiV$_2$O$_4$}


\author{R.\ Arita$^1$, K.\ Held$^2$, A.\ V.\ Lukoyanov$^{3}$ 
and V.\ I.\ Anisimov$^{4}$}
\affiliation{$^1$RIKEN (The Institute of Physical and Chemical Research), 
Wako, Saitama 351-0198, Japan\\
$^2$ Max-Planck-Institut f\"ur
Festk\"orperforschung, 70569 Stuttgart, Germany\\
$^3$ Ural State Technical University-UPI, 620002 Yekaterinburg, Russia\\
$^4$ Institute of Metal Physics, Russian Academy of Science-Ural division,
620219 Yekaterinburg, Russia
}

\date{\today}

\begin{abstract}
 We investigate the 
electronic structure of LiV$_2$O$_4$, for which
 heavy fermion behavior has been observed in various experiments, by the combination of the local density approximation 
and  dynamical mean field theory.
To obtain results at zero temperature,
we employ the projective quantum Monte Carlo method
as an impurity solver. 
Our results show that the strongly correlated $a_{1g}$ band is a lightly doped Mott insulator which -at low temperatures- shows a sharp (heavy) quasiparticle peak just above the Fermi level,
 which is
consistent with recent photoemission experiment by
Shimoyamada {\it et al.} [Phys. Rev. Lett. {\bf 96} 026403 (2006)].
\end{abstract}

\maketitle

The discovery of heavy fermion (HF) behavior in
the $3d$ material LiV$_2$O$_4$ 
\cite{LiVOexpKondo} 
was a big surprise since this phenomenon
was previously a hallmark of certain $f$ electron  
compounds. But below a characteristic temperature 
$T_K\sim 28\,$K,
the  linear specific heat coefficient ($\gamma$) \cite{LiVOexpKondo}, the
magnetic susceptibility \cite{LiVOexpKondo},the  Gr\"uneisen parameter \cite{Kondo97},  and the
 quadratic resistivity coefficient \cite{Urano2000}
 are also
for  LiV$_2$O$_4$
extraordinarily large, similar to
$f$ electron HF compounds and
much larger than in other transition metal oxides. From 
$\gamma\sim 420\,$mJ/molK$^2$, 
an effective mass enhancement of $m^*/m\sim 25$
was inferred \cite{Matsuno99}.
Neutron scattering \cite{neutron}, nuclear magnetic resonance \cite{NMR} and electron spin resonance \cite{ESR}, as well
as muon spin relaxation experiments \cite{muon}
indicate the existence of local magnetic moments,
which is consistent with a Curie-Weiss  susceptibility \cite{LiVOexpKondo}  in the temperature range 50 to 1000 K.
Down to the lowest temperatures measured  LiV$_2$O$_4$ remains a cubic spinel and 
 no long-range
magnetic, spin glass,   or superconducting order was 
observed.
More recently, also a sharp  Kondoesque peak of width 10$\,$meV
was observed in photoemission experiments \cite{Shimoyamada06} just  4 meV
above the
Fermi energy $E_F$, with a strong temperature dependence similar to
that of other HF compounds.
This finding is supported by the measurement of the magnetic curves
at $T=1.3$ K \cite{Niitaka2006} which also suggests the existence of a sharp peak 
slightly above $E_F$.

The  explanation of the
 HF behavior in  LiV$_2$O$_4$
has been a challenge since its discovery.
Local density approximation (LDA) calculations \cite{Matsuno99,Anisimov,Eyert} show
a twofold-degenerate and 2$\,$eV-wide
$e_g^\pi$ and a nondegenerate $a_{1g}$ band of width 1$\,$eV
cross the Fermi energy, 
 filled  altogether with 1.5 electrons per V ion.
This LDA bandstructure led one of the present
authors (VIA) 
to the proposal \cite{Anisimov,Kusunose} that
the $a_{1g}$ electrons play the role of the localized
 $f$ electrons in conventional HF compounds
 and the  $e_{g}^\pi$ electrons 
that of the itinerant valence electrons.
On the other hand, the importance of geometrical 
frustration originating in the spinel 
structure
has been stressed by various authors \cite{Lacroix,Shannon,Fulde,Burdin,Hopkinson,Fujimoto,Tsunetsugu,Yamashita,Laad}, 
all suggesting different 
explanations for the mass enhancement
of   LiV$_2$O$_4$.
Naturally the geometrical frustration suppresses any kind of long range 
order, so that local spin or orbital fluctuations 
should be dominant
as  suggested in Ref.\ \cite{Tsunetsugu,Yamashita}.

In this situation, we might  expect dynamical mean field
theory  (DMFT) \cite{DMFT} 
to be  good approximation for studying
the electronic correlations in this material. 
Realistic LDA+DMFT \cite{LDADMFT} calculations for  LiV$_2$O$_4$
have been carried out before
\cite{Nekrasov2003}, but neglected the $a_{1g}$-$e_{g}^\pi$ 
hybridization, which should be the driving force for the HF behavior in the Kondo  scenario \cite{Anisimov} and
were  furthermore restricted to temperatures $T> 750\,$K, far above $T_K$. Not surprisingly, a quasiparticle resonance was not found 
and the competition between 
antiferromagnetic direct exchange from
the $a_{1g}$-$a_{1g}$ hybridization,
ferromagnetic double exchange
from the $e_g^\pi$-$e_g^\pi$
hybridization, and the Kondo effect from the
(neglected) 
$a_{1g}$-$e_g^\pi$ hybridization
left this LDA+DMFT study \cite{Nekrasov2003} inconclusive.
Since that time the more sophisticated 
 projection onto Wannier functions has been 
developed \cite{Projection} which properly takes the
orbital off-diagonal hybridization into account.
Also the problem that conventional quantum Monte Carlo  (QMC)
simulations \cite{HFQMC} of the DMFT impurity problem
were restricted to rather high temperatures
because the numerical effort is proportional to $1/T^3$
has been overcome by the projective QMC 
(PQMC)
method \cite{Feldbacher,Arita,Arita2} for $T=0$.

With these improvements, we reinvestigate  LiV$_2$O$_4$
by LDA+DMFT(PQMC) and solve the puzzle why
this material shows HF behavior with a sharp 
Kondoesque resonance above the Fermi level.

\begin{figure}
\begin{center}
\includegraphics[width=8.0cm]{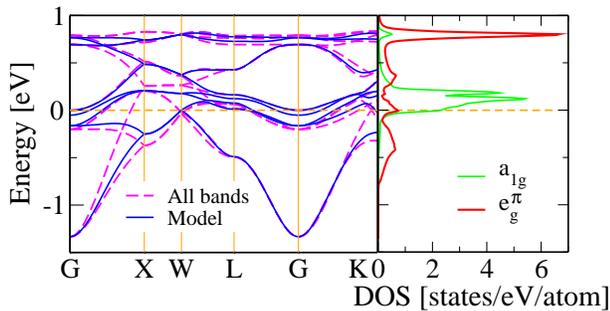}
\end{center}
\caption{(Color online) Left panel: Band dispersion of the effective 2-orbital model
(solid line) and total band structure (dashed line) of LiV$_2$O$_4$. 
Right panel: partial $a_{1g}$ and $e_g^\pi$ 
DOS for the model. $E_F$ is
 set to zero.
}
\label{Fig:Band}
\end{figure}

{\it Method.}
The unit cell of LiV$_2$O$_4$ contains four V atoms
and each V atom has three $t_{2g}$ orbitals,
which are split into the $a_{1g}$ orbital
and two degenerate $e_{g}^\pi$ orbitals due to the trigonal splitting.
First, we do a LDA calculation for LiV$_2$O$_4$ 
using the linearized muffin tin orbital basis set \cite{LMTO}. From this we further construct an effective 12 by 12 Hamiltonian 
by the projection onto Wannier functions \cite{Projection}. 
Since the $e_{g}^\pi$ orbitals are degenerate, it is possible 
to derive a 2-orbital model with an 8 by 8 Hamiltonian by taking 
only one of two $e_g^\pi$ orbitals into account. This
drastically decreases the 
computational efforts of the LDA+DMFT calculation
and hence allows for more accurate data.
As we will see {\em a posteriori}, the restriction to one 
 $e_g^\pi$ orbital will be justified by the fact that the
 $e_g^\pi$ orbitals play a rather passive role and the physics is determined by the $a_{1g}$ band.
The comparison of the band 
dispersion of this simplified 8-band model with 
total LDA
band structure of LiV$_2$O$_4$ in Fig.\ \ref{Fig:Band} shows
that
 the 2-orbital simplification  captures 
the essential features of the real compound's band 
structure. It also gives the densities of states (DOS) 
close to that previously reported by LDA, see Fig.\ 5 in \cite{Nekrasov2003}. 

Second, we supplement 
 this 2-orbital Hamiltonian 
by local intra- ($U$) and inter-orbital ($U'$) 
Coulomb repulsions as well as by Hund's exchange ($J$),
and solve the constructed many-body model by DMFT. It should be noted that we explicitly 
consider the off-diagonal elements between $e_g^\pi$ and $a_{1g}$ 
in contrast to all previous calculations \cite{Nekrasov2003}, 
where only the initial LDA $a_{1g}$-$e_g^\pi$ hybridization 
 is  reflected  indirectly in the DOS. This
$a_{1g}$-$e_g^\pi$
hybridization is essential 
for the Kondo effect with localized $a_{1g}$ and itinerant 
$e_g^\pi$ electrons \cite{Anisimov}.
As for the self-energy, we only consider the
diagonal element, so that the effective DMFT
impurity model becomes a two-orbital problem.
We also assume that the Hund coupling is of Ising type
since simulating 
the SU(2) symmetric Hund coupling is difficult 
in QMC. The application of new, more
sophisticated algorithms to this end, such as \cite{Sakai},
remains an important challenge for the future.

Besides conventional QMC, we employed the PQMC method in the present
LDA+DMFT calculation, basically following
Ref.\ \cite{Feldbacher,Arita,Arita2} for
calculating ground  state expectation values.
With 
an imaginary time discretization
$\Delta\tau=0.267 {\rm eV}^{-1}$, we
take ${\cal L}=20$ time slices for measurement
and ${\cal P}=65$ time slices before and thereafter
for projection.
For the  remaining imaginary time to  $\tilde{\beta}= \infty$,
we use the non-interacting Hamiltonian with a shifted
one-particle potential so that we have 
 $n=1$ electrons/site for the $a_{1g}$  orbitals and 
$n=0.25$ for the $e_{g}^{\pi}$ orbitals.
This shift 
warranties (approximately) the 
same large-$\tau$ asymptotic behavior as the 
interacting Hamiltonian.
We performed $\sim 3\times 10^8$ QMC sweeps
and used the maximum entropy
method for calculating the
spectral function  $A(\omega)$ and
the Fourier transformation of the Green
function from imaginary time to  frequencies,
i.e., from $G(\tau)$ to $G(i\omega)$.

\begin{figure}
\begin{center}
\includegraphics[width=6.5cm]{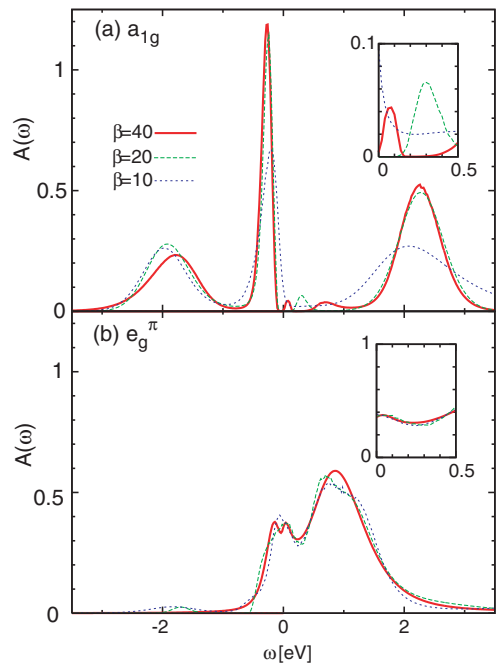}
\end{center}
\vspace{-.2cm}

\caption{(Color online)
Spectral function of LiV$_2$O$_4$ at $\beta=10,20,40$ 
eV$^{-1}$ ($T\approx1200, 600$ and 300 K);
$U=3.6$, $U'=2.4$, and $J=0.6$ (all units are in eV).
The insets show a closeup view  immediately
above  $E_F$ ($\omega=0$).
}
\label{Fig:A-FiniteT}
\end{figure}

{\it Results.}
Let us start with the results of conventional QMC at finite $T$.
In Fig.\ \ref{Fig:A-FiniteT}, we plot the spectral 
function $A(\omega)$, using Coulomb interaction parameters
which are typical for $3d$ orbitals, i.e., $U=3.6$, $U'=2.4$,
and $J=0.6$ eV. 
The qualitative feature of the 
result for $\beta\equiv 1/T=10$ eV$^{-1}$ corresponding 
to $T\approx1200$ K is similar to that of the previous
LDA+DMFT calculation \cite{Nekrasov2003}. 
But for $\beta=40$ eV$^{-1}$
($T\approx300$ K) we note the emergence of a small structure 
in the $a_{1g}$ band just above  $E_F$, which is absent for $T\approx1200$ K.
At the same time, we see no noteworthy temperature dependence for 
the $e_g^{\pi}$ band, especially around $E_F$ (see the inset). 
As far as the finite-$T$ QMC is concerned,
it is not clear whether the small structure
in the $a_{1g}$ band  becomes 
a sharp quasiparticle peak at lower temperatures.

\begin{figure}
\begin{center}
\includegraphics[width=6.3cm]{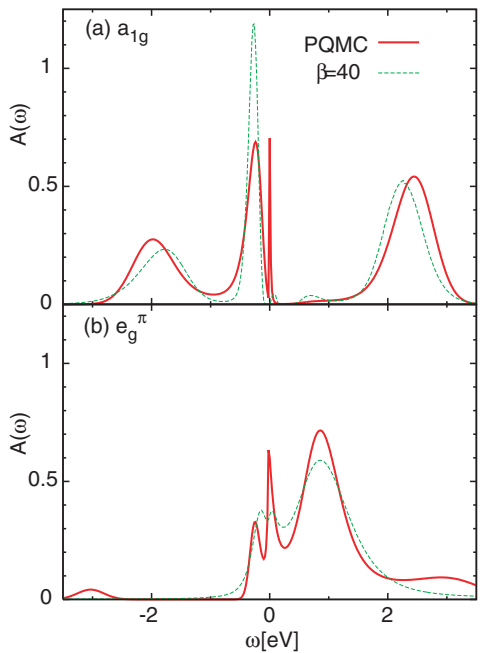}
\end{center}

\vspace{-.2cm}

\caption{(Color online)
Same as Fig.\ \ref{Fig:A-FiniteT} but now at $T=0$ (PQMC), compared to
$\beta=40$ eV$^{-1}$ corresponding to $T\approx300$ K.
In agreement with experiment, we can see a sharp peak slightly above $E_F$ (set to zero)
in the $a_{1g}$ band.
}
\label{Fig:PQMC-J06-A}
\end{figure}

\begin{figure}
\begin{center}
\includegraphics[width=6.2cm]{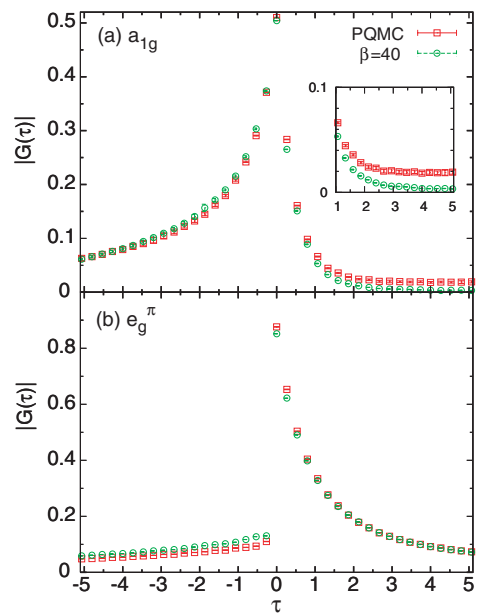}
\end{center}

\vspace{-.2cm}

\caption{(Color online)
Green function  for the $a_{1g}$ (a) and
$e_{g}^\pi$ (b) orbitals obtained by LDA+DMFT(PQMC)
as a function of $\tau$, compared with
the result for $\beta=40$ eV$^{-1}$ ($T\approx300$ K). 
In the inset, we magnify the region $\tau>1$. 
}
\label{Fig:PQMC-J06-G}
\end{figure}

To clarify this point, let us now turn to the 
PQMC results at $T=0$, see  Fig.\ \ref{Fig:PQMC-J06-A}.
Indeed we can see in the PQMC spectrum
that the small structure just above 
$E_F$ at $T\approx300$ K becomes a sharp peak, 
which is consistent with the experiment \cite{Shimoyamada06},
i.e., a peak 4 meV above $E_F$ whose width is 10 meV.
This $a_{1g}$ band is lightly doped, containing
 $n = 0.98$ electrons/site. Unfortunately,
the exact determination of the 
renormalization factor ($Z$) from
$A(\omega)$ or the self energy is difficult
because of the smallness of the structure and fluctuations
 from iteration
to iteration in the DMFT cycle. However,
the peak itself is stable as is the behavior of
 $G(\tau)$, plotted in Fig.\ \ref{Fig:PQMC-J06-G}.
The latter shows a  very slow decay
 for large $\tau$ in  PQMC which necessitates
the existence of a  sharp peak at small positive energies
in the $a_{1g}$ band.
In contrast, for $T\approx300$ K,  $G(\tau)$
vanishes exponentially, see  Fig.\ref{Fig:PQMC-J06-G} (a).

{\it Discussion.}
Let us now turn to the physical origin of the sharp peak in the $a_{1g}$ band.
One possible scenario which was originally proposed in 
Ref.\ \cite{Anisimov} is the Kondo effect caused 
by the hybridization between $a_{1g}$ and $e_{g}^\pi$ orbitals on neighboring sites (note that the onsite
hybridization is absent). However, it is not trivial whether
the associated (antiferromagnetic)
Kondo coupling is strong enough to survive 
 a Hund's exchange coupling as large as $J=0.6$.

\begin{figure}[h]
\begin{center}
\includegraphics[width=6.4cm]{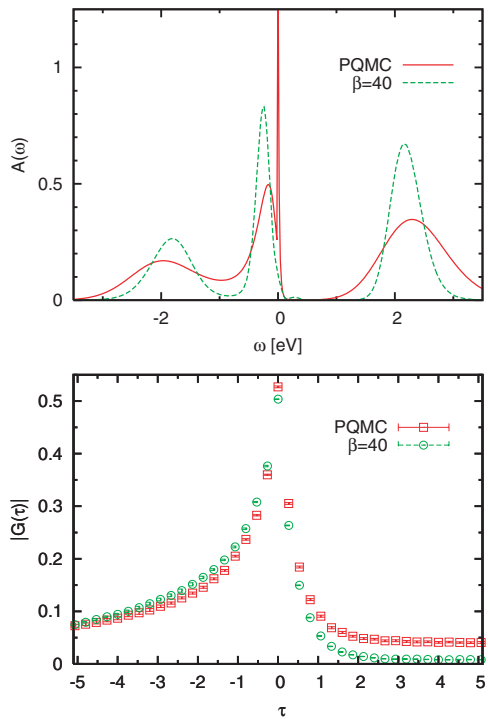}
\end{center}
\vspace{-.3cm}

\caption{
Spectrum (top) and
Green function (bottom)  of the $a_{1g}$ band
without $a_{1g}$-$e_g^\pi$ hybridization.
The sharp peak above $E_F$ 
survives if the hybridization is switched off.
}
\label{Fig:Fig6}
\end{figure}

To single out the effect of the
 $a_{1g}$-$e_{g}^\pi$
 hybridization,
we perform an auxiliary LDA+DMFT calculation.
To this end, we first  obtain   
the $a_{1g}$ and $e_{g}^\pi$
LDA DOS  from 
the effective 2-orbital Hamiltonian and
then do DMFT calculations with these DOSes
without any  hybridization.
We plot the resulting $G(\tau)$ and $A(\omega)$
of the $a_{1g}$ band in Fig.\ \ref{Fig:Fig6}.
Clearly, the sharp peak just above $E_F$ survives
switching off the $a_{1g}$-$e_{g}^\pi$
 hybridization. Hence, we can conclude that
 the Kondo scenario due to the hybridization
with $e_{g}^\pi$ orbitals 
\cite{Anisimov} cannot be
the microscopic origin for the peak in the $a_{1g}$ band. 
Actually, besides contributing to the doping of the  $a_{1g}$
band, the $e_{g}^\pi$ electrons 
do not play a pronounced role 
and are only weakly correlated.
Their self energy (not shown) is almost constant 
down to
very low frequencies $\sim 0.01\,$eV.
The constant (${\rm Im} \Sigma \sim - 0.14\,$eV) 
can be explained
by non-interacting electrons scattering at disordered
spins which is an appropriate description of the
 $a_{1g}$ electrons except for the lowest energies.

Hence, let us turn to 
 the
 $a_{1g}$ band itself which is not exactly half-filled, but
lightly doped with $n\sim 0.98$ electrons/site.
This suggests that the $a_{1g}$ band is a
 lightly-hole-doped Mott insulator
 with a  very strongly renormalized
quasiparticle because of the
nearness to the doping-controlled
Mott-Hubbard transition.
We can compare our results
with those of Ref.\ \cite{Pruschke}
for the single-band Hubbard model on the
 hypercubic lattice.
At $n=0.97$
 and very low temperature,
these results show
 a sharp peak just above $E_F$ \cite{Pruschke}, very similar to
our LDA+DMFT calculations.
An important question for this scenario of a doped Mott insulator
is whether the strong renormalizations
can survive the presence of short-range correlations beyond DMFT. In this respect, 
 the correlator projection method
indicates that $Z$ does not vanish for the filling-control 
Mott-Hubbard transition in the two-dimensional
 Hubbard model \cite{Imada},
and  also the dynamical vertex approximations \cite{Dvertex}
shows a strong damping of the quasiparticle peak 
in the vicinity of the Mott-Hubbard transition
due to antiferromagnetic fluctuations  beyond-DMFT.
However, we believe that such effects are less relevant 
for LiV$_2$O$_4$ because of the frustrated three-dimensional lattice and because there is no indication that the system 
is close to a magnetic 
phase transition.

{\it In conclusion,}
realistic LDA+DMFT calculations
for  LiV$_2$O$_4$
show a sharp peak for  $T \rightarrow 0$
in agreement with photoemission 
experiments and large renormalizations of the effective mass.
The physical origin of this peak is the 
lightly doping of the $a_{1g}$ band
which is hence metallic
but very close to a Mott-Hubbard transition.
The HF physics is not caused by
the hybridization between localized $a_{1g}$
and itinerant  $e_g^\pi$ orbitals.
Instead the  $a_{1g}$ orbitals play both roles simultaneously,
whereas the $e_g^\pi$ orbitals
are rather passive and not strongly correlated.


We would like to thank A. Georges, M. Imada, J. Matsuno, and S. Niitaka
for useful comments and 
acknowledge support by the Emmy Noether program of 
the Deutsche Forschungsgemeinschaft (KH), by
projects no. 06-02-81017, 04-02-16096, and 03-02-39024
of the Russian Foundation  
for Basic Research (AVL,VIA),
and by the Dynasty Foundation  (AVL).
Numerical calculations were done at the Supercomputer Center, 
Institute for Solid State Physics, University of Tokyo.


\begin{references}



\bibitem{LiVOexpKondo} S. Kondo {\it et al.}, 
Phys. Rev. Lett. {\bf 78}, 3729 (1997).

\bibitem{Kondo97} O.~Chmaissem {\it et al.}, 
Phys. Rev. Lett. {\bf 79}, 4866 (1997).

\bibitem{Urano2000} C. Urano {\it et al.}, 
Phys. Rev. Lett. {\bf 85}, 1052 (2000).


\bibitem{Matsuno99}J.~Matsuno, A.~Fujimori, and L.F.~Mattheiss,
Phys. Rev. B \textbf{60}, 1607 (1999).


\bibitem{NMR} N.~Fujiwara, H.~Yasuoka, and Y.~Ueda, 
Phys. Rev. B {\bf 57}, 3539 (1998);
A.\ V.~Mahajan {\em et al.},
Phys. Rev. B {\bf 57}, 8890 (2000).


\bibitem{neutron}
 A.~Krimmel {\em et al.},
Phys. Rev. Lett. {\bf 82}, 2919 (1999);
Phys. Rev. B {\bf 61}, 12578 (1998);
 S.-H.~Lee {\em et al.},
Phys. Rev. Lett. {\bf 86}, 5554 (2001).


\bibitem{ESR} M.~Lohmann {\em et al.},
Physica B {\bf 259-261}, 963 (1999).

\bibitem{muon} A. Koda {\em et al.}
Phys. Rev. B {\bf 69}, 012402 (2004).

\bibitem{Shimoyamada06}
A. Shimoyamada {\it et al.},
Phys. Rev. Lett. {\bf 96}, 026403 (2006).

\bibitem{Niitaka2006}
S. Niitaka {\it et al.}, presentation at 
ICM Kyoto (2006).


\bibitem{Anisimov}
V. I. Anisimov {\it et al.}, 
Phys. Rev. Lett. {\bf 83}, 364 (1999).

\bibitem{Eyert}
V. Eyert {\it et al.},  
Europhys. Lett. {\bf 46}, 762 (1999).


\bibitem{Kusunose} Also note
H. Kusunose, S. Yotsuhashi and K. Miyake, 
Phys. Rev. B {\bf 62}, 4403 (2003).

\bibitem{Lacroix} 
C. Lacroix, Can. J. Phys. {\bf 79}, 1469 (2001).

\bibitem{Shannon} 
N. Shannon, Eur. Phys. J. B {\bf 27}, 527 (2001).

\bibitem{Fulde}
P. Fulde  {\it et al.}, 
Europhys. Lett. {\bf 54}, 779 (2001).

\bibitem{Burdin}
S. Burdin, D. R. Grempel, and A. Georges,
Phys. Rev. B {\bf 66}, 045111 (2002).

\bibitem{Hopkinson} 
J. Hopkinson and P. Coleman, 
Phys. Rev. Lett. {\bf 89}, 267201 (2002).

\bibitem{Fujimoto} 
S. Fujimoto, Phys. Rev. B {\bf 65}, 155108 (2002).



\bibitem{Tsunetsugu}
H. Tsunetsugu, J. Phys. Soc. Jpn {\bf 71}, 1845 (2002).

\bibitem{Yamashita}
Y. Yamashita and K. Ueda, Phys. Rev. B {\bf 67}, 195107 (2003).

\bibitem{Laad} 
M. S. Laad, L. Craco, and E. M{\"u}ller-Hartmann,
Phys. Rev. B, {\bf 67}, 033105 (2003).

\bibitem{DMFT} 
W. Metzner and D. Vollhardt,
Phys. Rev. Lett. {\bf 62},  324  (1989);
A. Georges and G. Kotliar,
Phys. Rev. B, {\bf 45},  6479 (1992);
A. Georges {\em et al.}, Rev. Mod. Phys. {\bf  68}, 13 (1996).



\bibitem{LDADMFT}
V.\ I.\ Anisimov {\it et al.}, J. Phys. Condens. Matter
{\bf 9}, 7359 (1997), A.\ I.\ Lichtenstein and M.\ I.\ Katsnelson,
Phys. Rev. B {\bf 57}, 6884 (1998);
K.\ Held {\it et al.}, phys. stat. sol. (b) {\bf 243}, 2599 (2006);  cond-mat/0511293; G. Kotliar {\em et al.}, Rev. Mod. Phys. {\bf 78}, 865 (2006).

\bibitem{Nekrasov2003}
I.A. Nekrasov {\it et al.}, 
Phys. Rev. B. {\bf 67}, 085111 (2003).


\bibitem{Projection} 
V. I. Anisimov {\it et al.}, 
Phys. Rev. B {\bf 71}, 125119 (2005).

\bibitem{HFQMC}
J.\ E.\ Hirsch and R.\ M.\ Fye, Phys. Rev. Lett.
{\bf 56}, 2521 (1986).

\bibitem{Feldbacher} M. Feldbacher, K. Held, and F. F. Assaad,
Phys. Rev. Lett. {\bf 93}, 136405 (2004).

\bibitem{Arita}
R. Arita and K. Held,
Phys. Rev. B {\bf 72}, 201102 (2005).

\bibitem{Arita2}
R. Arita and K. Held,
Phys. Rev. B {\bf 73}, 064515 (2006).

\bibitem{LMTO} O. K.~Andersen, Phys. Rev. B {\bf 12}, 3060 (1975);
O.~Gunnarsson, O.~Jepsen, and O. K.~Andersen, Phys. Rev. B {\bf 27}, 7144 (1983).

\bibitem{Sakai}  S. Sakai {\it et al.}, 
Phys. Rev. B. {\bf 74}, 155102 (2006).

\bibitem{Pruschke} Th. Pruschke, D. L. Cox, M. Jarrell,
Phys. Rev. B. {\bf 47}, 3553 (1993).

\bibitem{Imada} K. Hanasaki and M. Imada, J. Phys. Soc. Jpn.
{\bf 75}, 084702 (2006).

\bibitem{Dvertex} A. Toschi, A. Katanin and K. Held,
cond-mat/0603100.
\end{references}
\end{document}